\documentclass[11pt,twoside]{article}
\usepackage{./asp2014}

\aspSuppressVolSlug
\resetcounters

\begin{document}

\title{Detecting Infall in High-Mass Protostellar Objects}
\author{Maria T.\ Beltr\'an$^1$, Qizhou Zhang$^2$, and V\'{\i}ctor M.\ Rivilla$^1$
\affil{$^1$INAF-Osservatorio Astrofisico di Arcetri, Largo E.\ Fermi 5,
I-50125 Firenze, Italy; \email{mbeltran@arcetri.astro.it, rivilla@arcetri.astro.it}}
\affil{$^2$Harvard-Smithsonian Center for Astrophysics, 60 Garden St. Cambridge, MA 02138, USA; \email{qzhang@cfa.harvard.edu}}
}

\paperauthor{Maria T.\ Beltr\'an}{mbeltran@arcetri.astro.it}{0000-0003-3315-5626}{INAF}{Osservatorio Astrofisico di Arcetri}{Firenze}{}{50125}{Italy}
\paperauthor{Qizhou Zhang}{qzhang@cfa.harvard.edu}{}{Harvard-Smithsonian Center for Astrophysics}{}{Cambridge}{MA}{02138}{USA}
\paperauthor{V\'{\i}ctor M.\ Rivilla}{rivilla@arcetri.astro.it}{0000-0002-2887-5859}{INAF}{Osservatorio Astrofisico di Arcetri}{Firenze}{}{50125}{Italy}


\section{Introduction: the formation of high-mass stars}

The role of accretion disks in the formation of low-mass stars has been well assessed by means of high angular resolution observations at various wavelengths.  These findings confirm the prediction that conservation of angular momentum during the collapse leading to the formation of a star is bound to produce
flattening and rotation of the collapsing core. Therefore, the existence of such disks around low-mass young stellar objects (YSOs)
strongly supports the scenario where the formation of {\it low-mass} stars proceeds through contraction of a dense molecular clump
and subsequent inside-out collapse with accretion onto a protostellar core. In synthesis, one can say that {\it low-mass stars form
through accretion}.

What about {\it high-mass} stars? The formation process of high-mass stars has puzzled the astrophysical community for decades because of the apparent stellar mass limit for spherical accretion. Beyond this limit, theory predicted that it is impossible to continue accreting material because the stellar wind and the radiation pressure from the newly-formed early-type star would stop the infall \citep[e.g.,][]{kahn74, yorke77, wolfire87}. In the 1990s, different theoretical scenarios proposed non-spherical accretion as a possible solution for the formation of OB-type stars \citep{nakano89, jijina96}, and in recent years, theoretical ideas and simulations appear to have converged to a disk-mediated accretion scenario \citep[e.g.,][]{krumholz09, kuiper10, klassen16}.  In fact, competing theories that propose very different high-mass star-formation mechanisms, such as models suggesting that massive star-formation is initiated by the monolithic collapse of a turbulent molecular core \citep{mckee02}, or those based on competitive accretion \citep{bonnell06}, all predict the existence of circumstellar accretion disks through which the material is channeled onto the forming star. With this in mind, it is clear that  the detection of infall in circumstellar disks represents a crucial test to prove the accretion scenario also applies in the case of high-mass stars.

At present, several authors have reported on detections of disks around high-mass YSOs \citep[see][for a review]{beltran16}. In most cases, the existence of such a disk is also supported by the detection of a bipolar jet/outflow along the apparent rotational axis of the disk. It must be noted that some of these disks resemble massive toroids rotating about a cluster of high-mass stars, rather than geometrically thin circumstellar disks. In any case, no matter whether the disk is circumstellar or ``circumcluster'', these findings strongly support the accretion scenario also for the high-mass regime. 

Notwithstanding these important results, the presence of disks rotating about high-mass stars is not sufficient by itself to prove unambiguously the accretion model: what is needed is iron-clad evidence of {\it infall}. Such evidence is very difficult to find, as the free-fall velocity becomes significant only very close to the accreting star, i.e., over a region of a few 0.01~pc ($\sim$2000\,au), which is very difficult to access and disentangle from the surrounding quiescent or rotating material.  In this chapter we discuss how to characterize the infall of material in a sample of 36 high-mass accretion disk candidates covering a broad range of luminosities, from 10$^3$\,$L_\odot$ to 10$^6$\,$L_\odot$, compiled by \citet{beltran16} with the ngVLA. 

\section{Tracing infall}

Red-shifted molecular absorption against a bright background continuum source is possibly the clearest way of diagnosing infall and accretion. For this, it is necessary to observe high-density tracers with excitation energies below the brightness temperature of the central continuum source, and observe them especially with interferometers (because the emission does not suffer from beam dilution). The first studies using this technique were conducted at centimeter wavelengths with the VLA by observing  NH$_3$ in absorption against the bright ultracompact (UC) H{\sc ii} region G10.62$-$0.4 \citep{ho86,keto88}.  Later one, infall has been detected toward other ultracompact/hypercompact (UC/HC) H{\sc ii} regions:  e.g., W51 \citep{zhang97}, G24.78+0.08 \citep{beltran06a}, NGC 7538\,IRS1 \citep{goddi15}, also observed in NH$_3$ with the VLA or the upgraded Jansky VLA. Ammonia is the best species to study infall in dense cores because  it is possible to trace excitation up to temperatures of $\sim$2000\,K by observing
its inversion transitions within a relatively narrow frequency range at centimeter wavelengths, 20--40\,GHz, which cannot be observed with (sub)millimeter interferometers such as SMA, IRAM NOEMA, or ALMA. This allows us to conduct the tomography of the infall in the core.  In addition, in case the main line is optically thick, one can always use the hyperfine satellite lines, which are usually optically thin, to study the absorption.

Red-shifted absorption at high-angular resolution has also been observed against bright dust continuum sour\-ces at millimeter wavelengths  \citep[e.g.,][]{zapata08, girart09, beltran18}. For high-mass protostars, however, dust emission from envelopes and disks becomes so optically thick that  deep probes of infall motion are prevented. In addition, the brightness temperatures of the UC/HC H{\sc ii} regions are typically 10$^3$--10$^4$\,K, much higher than those of the dust continuum sources (a few hundreds of K). This bright background enables detection of line absorption with much higher signal-to-noise ratio, as well as probing infall closer to the central protostar with higher energy molecular line transitions. 

{\it What can we learn from the gravitational collapse by observing red-shifted absorption?} We can estimate the infall rate $\dot M_{\rm inf}$ of material in the core. Note that red-shifted absorption profiles trace infalling material in the inner part of the core or in the disk, but not actual accretion onto the central star. Following \cite[][]{beltran06a}, the infall rate  in a solid angle $\Omega$ can be estimated from the expression $\dot M_{\rm inf} = (\Omega/4\,\pi)\,2\,\pi\,m_{\rm H_2}\,N\,R^{1/2}_0\,R^{1/2}\,V_{\rm inf}$,  where $N$ is the gas column density, $R_0$ is the radius of the continuum source, $V_{\rm inf}$ is
the infall velocity, and $R$ is the radius at which $V_{\rm inf}$  is measured. The main caveat of this method is the uncertainty on $R$. In fact, the radius at which $V_{\rm inf}$  is associated is not known, and usually the size of the infalling core is given as an upper limit. Red-shifted absorption can also help us to constrain star-formation theories by studying whether infall spins down or spins up. In fact, the advantage of  tracing a wide range of excitation temperatures ($\sim$20\,K to 2000\,K) with several transitions of NH$_3$ is that by measuring the velocity of the absorption feature for each transition it is possible to study whether the infall velocity (assumed to be the difference between the velocity of the red-shifted absorption dip and the systemic LSR velocity) changes with the energy of the line. If the infall velocity increases with the energy of the transition (spins up), this would suggest that the infall is accelerating towards the center of the core, consistent with gravitational collapse \citep{beltran18}.  One the other hand, if the infall velocity decreases (spins down), it could suggest that magnetic braking is acting and removing angular momentum \citep{basu94}. This would suggest that the gravitational collapse of the core is controlled by magnetic fields  \citep{girart09}. In addition, a detailed modeling of the NH$_3$ line emission and of the spectral energy distribution with an infalling envelope model can give us information on the mass of the central star, its age and luminosity, and the mass accretion rate, similar to what \citet[][]{osorio09} have done for the hot molecular core G31.41+0.31.

\section{Uniqueness to ngVLA capabilities}

Up to now, red-shifted absorption at centimeter wavelengths has been only detected toward the brightest UC/HC H{\sc ii} regions and with limited angular resolution and sensitivity using the VLA. These limits mean that, in many cases, the absorption has not been spatially resolved and only the lower excitation energy transitions, which are probably tracing the outer regions of the massive envelope, have been observed.
Therefore, to trace infall in high-mass cores, with luminosities ranging from  10$^3$\,$L_\odot$ up to 10$^6$\,$L_\odot$, high sensitivity and high-angular resolution are essential. High sensitivity is needed to trace high-excitation transitions of high-density tracers, which are those tracing the material closest to the protostar. On the other hand, high-angular resolution would allow us to resolve spatially the absorption, even for distant objects. A good tracer of infall is NH$_3$ because, as already mentioned, multi transitions at different energies can be observed in red-shifted absorption against bright ionized sources. 

\articlefigure{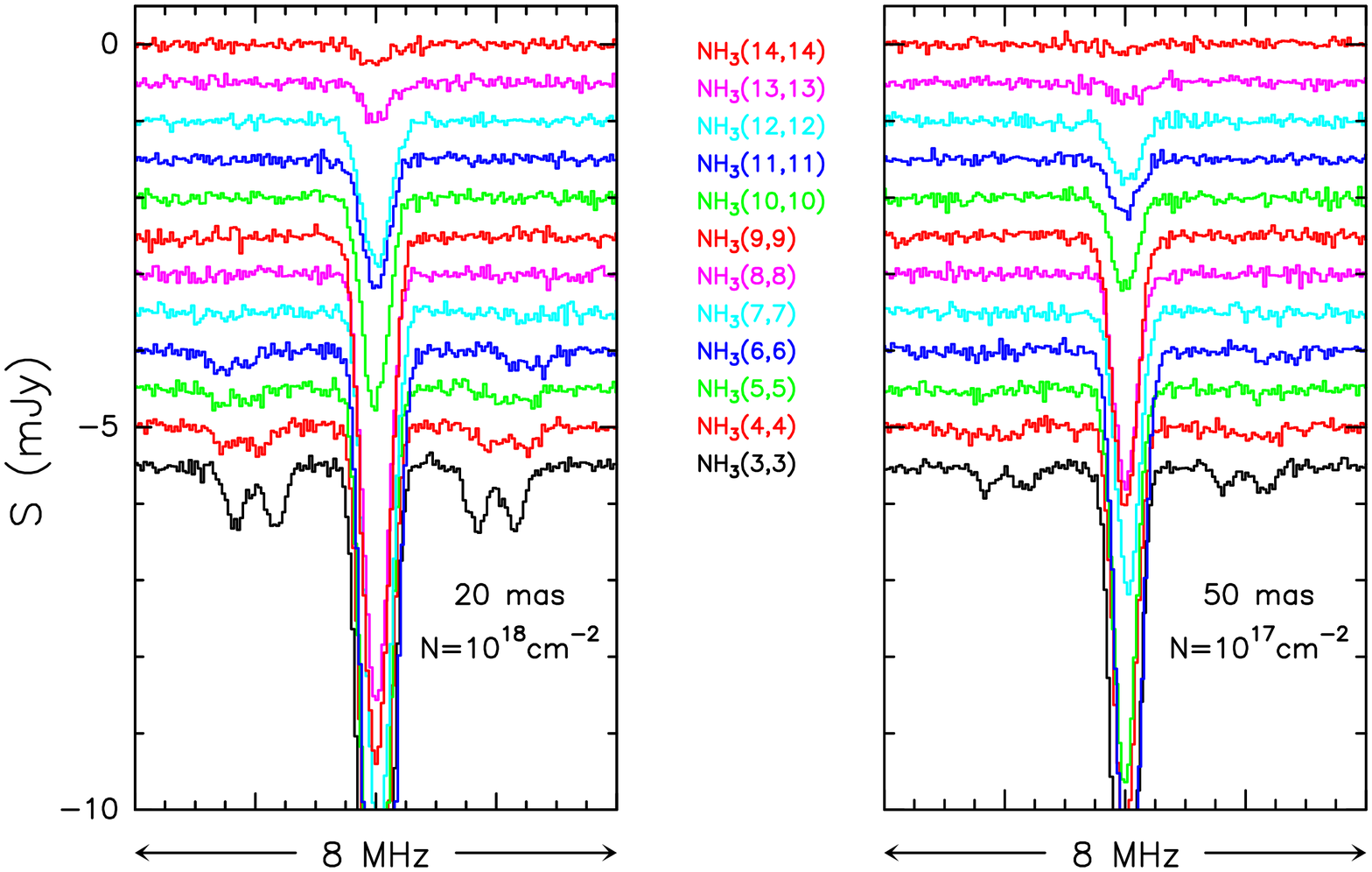}{fig:nh3}
{Synthetic spectrum of NH$_3$ obtained with MADCUBA, assuming LTE conditions.
Different transitions, from the (3,3) inversion transition to the (14,14) one, are shown with different colors. For display purposes, the transitions have been shifted by 0.5 mJy in the y-axis.
The left and right panels show two cases for different angular resolutions and column densities of NH$_3$: 20 mas and 10$^{18}$ cm$^{-2}$,  and 50 mas and 10$^{17}$ cm$^{-2}$, respectively. The linewidth considered is 5\,km\,s$^{-1}$. The brightness temperature of the UC/HC H{\sc ii} region is 10$^4$\,K. We have considered a noise level of 45 $\mu$Jy, which is the rms achieved for an integration time of 10 hr and a spectral resolution of 1 km s$^{-1}$, according to the specifications of the ngVLA.}


The high sensitivity of the ngVLA will allow us to detect NH$_3$ inversion transitions  in red-shifted absorption toward all the sources of the sample of high-mass accretion disk candidates with luminosities ranging from 10$^3$\,$L_\odot$ up to 10$^6$\,$L_\odot$ compiled by \citet{beltran16}, including the faintest ionized ones. In particular, the ngVLA will detect infall toward massive stars shortly after the onset of an H{\sc ii} region, and at a maximum angular resolution of 20\,mas. Figure~\ref{fig:nh3} shows the synthetic spectrum of NH$_3$,  from the (3,3) inversion transition to the (14,14) one, obtained with MADCUBA (Madrid Data Cube Analysis on ImageJ), which is a software developed in the Center of Astrobiology (Madrid, INTA-CSIC) to visualize and analyze single spectra and datacubes (\citealt{rivilla16}; Mart{\'i}n et al., in prep.). The spectra have been modeled for two different angular resolutions, 20 mas and 50 mas with  different column densities of NH$_3$, $10^{18}$\,cm$^{-2}$ and  $10^{17}$\,cm$^{-2}$, respectively. As seen in the figure, an rms of $\sim$45\,$\mu$Jy/beam, achievable after a 10 hr integration per source,  would allow us to detect the main lines of the weakest inversion transitions of ammonia, up to the NH$_3$ (14, 14) transition, and the satellite lines of the strongest ones, for NH$_3$ column densities consistent with (or even lower than) those found around high-mass young stellar cores \citep[e.g.,][]{beltran06a, goddi15}. Spectral resolution should not be a problem for the ngVLA,  considering that a modest channel spacing of $\sim$1\,km/s is less than the typical line widths of high-mass protostellar cores of a few km/s.  Regarding the angular resolution, 20\,mas corresponds to 100\,au at a distance of 5\,kpc, which is the typical distance of O-type young stars \citep[e.g.,][]{beltran06b}. Typical sizes of accretion disks around low-mass stars are of the same order, $\sim$100\,au. Note that the highest angular resolution provided by VLA at the NH$_3$ transition frequencies is about 50\,mas. Given the high mass of these cores, they often fragment and form multiple protostars. Therefore, the high-angular resolution of ngVLA will resolve the emission and trace the accretion flows around individual protostars.

 The milli-arcsecond angular resolution achieved with the ngVLA and the wide range of excitation energies covered by the NH$_3$ inversion transitions (from $\sim$20\,K to 2000\,K) will allow us to trace the kinematics of circumstellar gas from the envelope down to the circumstellar disk, and infer the properties of the accretion to better constrain theories of massive star formation. In fact, although the two competing theories on massive star-formation propose a disk-mediated accretion scenario, the properties of such disks are expected to be different. In fact, while in the turbulent core collapse case one would expect the material to be incorporated onto the central protostar though a true accretion disk in Keplerian motion with a size of a few hundreds of au, in the competitive accretion, the disks could be asymetric, truncated and very small ($<100$\,au) due to tidal interactions \citep[][]{cesaroni06} in the crowded star-forming cores (according to this model, high-mass stars always form in clusters). Alternatively, accretion could proceed to the central protostar through filaments rather than through a disk \citep[][]{goddi18}. 



\section{Synergy at other wavelengths}

The study of infall at centimeter wavelengths complements: {\it i)} ALMA or NOEMA  observations of molecular outflows aimed at estimating the mass outflow rate, which is related to the mass accretion rate onto the protostar as $\dot M_{\rm acc}=\dot M_{\rm out}/6$ \citep{tomisaka98, shu99}, 
and {\it ii)} JWST observations of Br$\gamma$ lines aimed at determining the mass accretion rate in the disk. By comparing the mass infall rate obtained at centimeter wavelengths to the mass accretion rate one could determine concretely whether the former are much higher than the latter, as suggested by recent studies \citep{beltran16}. Such a mismatch which could indicate pile-up of material in the disk and the existence of episodic accretion, for example as recently observed in the high-mass stellar object S255IR \citep{caratti17}.

\section{Conclusions}

The superb sensitivity and angular resolution provided by the ngVLA  will allow us to study the gravitational collapse in high-mass star-forming sources of all luminosities (from 10$^3$\,$L_\odot$ to 10$^6$\,$L_\odot$). The detection of red-shifted absorption in several inversion transitions of NH$_3$ with different excitation energies will allow us to trace the material down to disk scales and to derive the properties of the collapse and of the accretion disks that will help us to better constrain  theories of high-mass star formation.  In particular, by studying the infall velocity field, i.e., whether infall spins down or spins up, we will infer what controls the gravitational collapse in high-mass star-forming regions. 

\acknowledgements 
V.M.R. is funded by the European Union's Horizon 2020 research and innovation programme under the Marie Sk\l{}odowska-Curie grant agreement No 664931. 


\begin{thebibliography}{}

\bibitem[Basu \& Mouschovias(1994)]{basu94}
Basu, S.,  \& Mouschovias, T.\ Ch.\ 1994, ApJ, 432, 720

\bibitem[Beltr\' an  \& de Wit(2016)]{beltran16} 
 Beltr\' an, M.\ T., \& de Wit, W.\ J.\ 2016, A\&ARv, 24, 6

\bibitem[Beltr\' an et al.(2006a)]{beltran06a}
Beltr\' an, M.\ T., Cesaroni, R., Codella, C., Testi., L.\ et al.\ 2006a, Nature, 443, 427 

\bibitem[Beltr\' an et al.(2006b)]{beltran06b}
Beltr\' an, M.\ T.,  Brand, J., Cesaroni, R.\ et al.\ 2006b, A\&A, 447, 221

\bibitem[Beltr\' an et al.(2018)]{beltran18}
Beltr\' an, M.\ T., Cesaroni, R., Rivilla, V.\ M.\ et al.\ 2018, A\&A, arXiv:1803.05300

\bibitem[Bonnell \& Bate(2006)]{bonnell06}
Bonnell, I.\ A., \& Bate, M.\ R.\ 2006, MNRAS, 370, 488

\bibitem[Caratti o Garatti et al.(2017)]{caratti17}
Caratti o Garatti, A.,  Stecklum, B., Garcia L\'opez, R.\ et al.\ 2017, Nature Physics, 13, 276

\bibitem[Cesaroni et al.(2006)]{cesaroni06}
Cesaroni, R., Galli, D., Lodato, G.\ et al.\ 200, Nature, 444, 703

\bibitem[Girart et al.(2009)]{girart09}
Girart, J.\ M., Beltr\'an, M.\ T.,  Zhang, Q.\ et al.\ 2009, Science,
324, 1408

\bibitem[Goddi et al.(2015)]{goddi15}
Goddi, C., Zhang, Q., \& Moscadelli, L.\ 2015, A\&A, 573, A108

\bibitem[Goddi et al.(2018)]{goddi18}
Goddi, C., Ginsburg, A., Maud, L.\ et al.\ 2018, arXiv:1805.05364

\bibitem[Kahn(1974)]{kahn74}
Kahn, F.\ 1974, A\&A, 37, 149

\bibitem[Keto et al.(1988)]{keto88}
Keto, E.\ R.,  Ho, P.\ T.\ P., \& Haschick, A.\ D.\ 1988, ApJ, 324, 920

\bibitem[Klassen et al.(2016)]{klassen16}
Klassen, M., Pudritz, R.\ E., Kuiper, R., Peters, Th., Banerjee, R.\ 2016, ApJ, 823, 28

\bibitem[Krumholz et al.(2009)]{krumholz09}
Krumholz, M.\ R.,  Klein, R.\ I., McKee, C.\ F., Offner, S.\ S.\ R., \& Cunningham,
A.\ J.\ 2009, Science, 323, 754

\bibitem[Kuiper et al.(2010)]{kuiper10}
Kuiper, R., Klahr, H., Beuther, H., \& Henning, T.\  2010, ApJ, 722, 1556826, 161

\bibitem[Ho \& Haschick(1986)]{ho86}
Ho, P.\ T.\ P., \& Haschick, A.\ D.\ 1986, ApJ, 304, 501

\bibitem[Jijina \& Adams(1996)]{jijina96}
Jijina, J., \& Adams, F.\ C.\ 1996, ApJ, 462, 874

\bibitem[McKee \& Tan(2002)]{mckee02}
McKee, C.\ F., \& Tan, J.\ C.\ 2002, Nature, 416, 59

\bibitem[Nakano(1989)]{nakano89}
Nakano, T.\ 1989, ApJ, 345, 464

\bibitem[Osorio et al.(2009)]{osorio09}
Osorio, M., Anglada, G., Lizano, S., D'Alessio, P.\ 2009, ApJ, 694, 29

\bibitem[Rivilla et al.(2016)]{rivilla16}
Rivilla, V.\ M., Fontani, F., Beltr\'an, M.\ T.\ et al.\ 2016, ApJ, 

\bibitem[Shu et al.(1999)]{shu99}
Shu, F., Allen, A., Shang, H.\ et al.\ 1999, NATO Adv.\ Sci.\ Institutes Ser.\ C., Lada, C.\ \& Kylafis, N.\ 
(ed) NATO Advanced Science Institutes (ASI) Series C,  540, 193

\bibitem[Tomisaka (1998)]{tomisaka98}
Tomisaka, K.\ 1998, ApJL, 502, 163

\bibitem[Wolfire \& Casinelli(1987)]{wolfire87}
Wolfire, M, \& Cassinelli, J.\ 1987, ApJ, 319, 850

\bibitem[Yorke \& Kr\"ugel(1977)]{yorke77}
Yorke, H. W., \& Kr\"ugel, E.\ 1977, A\&A, 	54, 183

\bibitem[Zapata et al.(2008)]{zapata08}
Zapata, L., Palau, A., Ho, P.\ T.\ P.\ et al.\ 2008, A\&A, 479, L25

\bibitem[Zhang \& Ho(1997)]{zhang97}
Zhang, Q., \& Ho, P.\ T.\ P.\ 1997, ApJ, 488, 241

\end{thebibliography}


\end{document}